# Magnetization Dynamics of Fibonacci-Distorted Kagome Artificial Spin Ice

Ali Frotanpour, Justin Woods, Barry Farmer, Amrit P. Kaphle, Lance E. De Long,

Department of Physics and Astronomy, University of Kentucky, 505 Rose Street, Lexington,

Kentucky 40506-0055, USA

Loris Giovannini and Federico Montoncello, Dipartimento di Fisica e Scienze della Terra,

Università di Ferrara, Ferrara, I-44121, ITALY

Email: frotanpour@uky.edu

lance.delong@uky.edu

## Abstract

We present results of ferromagnetic resonance (FMR) experiments and micromagnetic simulations for a distorted, 2D Kagome artificial spin ice. The distorted structure is created by continuously modulating the 2D primitive lattice translation vectors of a periodic honeycomb lattice, according to an aperiodic Fibonacci sequence used to generate 1D quasicrystals. Experimental data and micromagnetic simulations show the Fibonacci distortion causes broadening and splitting of FMR modes into multiple branches, which accompany the increasing number of segment lengths and orientations that develop with increasing distortion. When the applied field is increased in the opposite direction to the net magnetization of a segment, spin wave modes appear, disappear or suddenly shift, to signal segment magnetization reversal events. These results show the complex behavior of reversal events, as well as well-defined frequencies and frequency-field slopes of FMR modes, can be precisely tuned by varying the severity of the aperiodic lattice distortion. This type of distorted structure could therefore provide a new tool for the design of complicated magnonic systems.

## Introduction

Recent interest in geometrical frustration of magnetic order has led to the fabrication of a variety of submicron, thin-film structures that can be systematically controlled using the tools of nanofabrication. Artificial spin ices (ASI) were originally comprised of elongated segments of magnetic thin-film deposited to form a 2D periodic lattice whose frustrated topology depressed long-range magnetic order [1]. The strong shape anisotropy (length l >> width w >> thickness t) of disconnected, sub-micron segments makes them behave as classical Ising spins in the "Ising-saturated" regime, which refers to a state in which every "macrospin" has a magnetization texture oriented parallel to its long axis (typically for applied fields below 1000 Oe in the case of magnetically soft $Ni_{0.8}Fe_{0.2}$ films).

Given the magnetization **M** of an isolated segment is largely uniform and parallel to the long axis, one can define opposite magnetic charges (~ div**M**) confined to either end of an isolated segment. These considerations allow the magnetic dipole interactions among Ising segments to be approximated by a "dumbbell" charge model [1] for binary Ising dipoles. The magnetostatic energy of an ASI can be calculated by enforcing a "spin ice rule" (SIR) that minimizes the total magnetic charge located near pattern vertices. This approach mimics the original development of the SIR that describes the ground state arrangement of atomic spins in the tetrahedral sublattice of pyrochlore spin ice [2], where frustration is generated among threefold tetrahedral bonds.

Therefore, ASI provides a 2D mesoscopic system in which lattice parameters and key magnetic interactions can be tuned by geometric design. Various techniques can also be used to characterize the magnetic state of ASI, including FMR and magnetotransport data. Moreover, nanoscale imaging techniques such as PEEM and SEMPA [1, 14] can be used to directly observe the magnetic state of ASI to better understand correlation effects and resulting phase transitions. Of particular interest are the honeycomb and related Kagome lattices, because they are composed of threefold vertices that promote strong frustration in the case of nearest-neighbor antiferromagnetic interactions. Over the last decade, researchers have explored the magnetization dynamics of Kagome ASI (KASI), and sought to confirm a predicted phase transition into long-range magnetic order [2-13].

However, the effects of subtle imperfections, disorder and reduced lattice symmetry on the stability of paramagnetism versus magnetic order in frustrated systems, together with what is sometimes referred to as "order out of disorder", have recently received increasing attention. In this context, a novel class of *aperiodic* lattices in the form of 2D artificial quasicrystals (AQC) has been introduced to explore effects of reduced lattice symmetry on magnetic order within frustrated sublattices, controlled magnetic switching, and novel spin dynamics [14-21].

It is important to keep in mind that ASI, as "metamaterials" (materials by design), offer new paradigms for development of locally manipulable magnetic memories, complex magnetic switching networks, and magnonic devices **[9, 15-17]**. In particular, spin waves in ASI are analogous to electric current in conducting networks in which charged information carriers can move at the cost of Joule heating; whereas spin waves can potentially move in a lossless fashion, which is promising for designs of spin-wave-operated logic gates **[22-23]**. Furthermore, Kagome ASI offer a simple symmetry for implementing "signal forking", which is a basic feature of neural networks; they therefore represent good candidates for creating structures for neuromorphic computing **[24-27]**. The introduction of the Fibonnaci distortion degree of freedom presents a novel adjustable parameter that can mimic the adaptability and plasticity of neural networks.

However, efforts to precisely control the "degree of disorder" presented by AQC are hampered by the fact that the exotic (e.g., fivefold or eightfold) rotational symmetries of AQC forbid their ***continuous distortion*** into a periodic Bravais lattice (i.e., 2D AQC are "topologically inequivalent" to periodic lattices). Nevertheless, it is possible to continuously distort a 2D Bravais lattice into an aperiodic array by utilizing an aperiodic ***Fibonacci sequence*** that is closely related to the structures of various types of quasicrystal arrays **[14-20]**. An example is shown in **Fig. 1**, where the spacings of incomplete planes of parallel segments in a Penrose P2 tiling follow a Fibonacci sequence of long (L) and short (S) distances. The fact that quasicrystals exhibit aperiodicity that can be precisely described by mathematical algorithms places them in a unique, intriguing category of ***controlled, intermediate disorder***, compared to ***random*** aperiodic arrays. We anticipate such intermediate systems will exhibit novel, interesting magnetic dynamics and ground state order **[14, 15]**, which is a prime motivation of the present study.

Herein, we report our FMR study of a connected Fibonacci-distorted KASI (FKASI), which permits the sixfold rotational and periodic translational symmetries of the undistorted honeycomb lattice to be continuously reduced; the resulting aperiodic lattice retains only mirror symmetry, and features modified shape anisotropies and magnetic moments of various film segments, as shown in **Fig. 2**. Most ASI studied to date have been ***disconnected***, periodic lattices of elongated film segments that mimic classical Ising dipoles **[2, 14, 15]**. However, magnetoresistive devices **[28-30]** and other potential applications of wire networks **[31, 32]** are more amenable to ***connected*** lattices of segments. However, the connections introduce complications from formation of magnetic domain walls (DW) and short-range exchange interactions within lattice vertices, which can modify SIR and alter magnetic ground states and magnetic reversal. Fortunately, systematic methods can be applied to take these complications into account [14]; and we will focus on connected KASI in this work.

Ferromagnetic resonance (FMR) spectroscopy is a proven, powerful probe of the magnetization dynamics of periodic KASI [7-10] and AQC [13, 15-18]. Combined with the interpretive help of micromagnetic simulations, FMR can detect resonances originating from a group of ASI segments with specific magnetization orientations with respect to the applied field [9]. For applied fields below the Ising saturation regime, FMR spectra largely reflect the orientation of Ising spins that belong to distinct sub-groups of segments. For example, a relatively high-frequency branch of FMR modes generally indicates a set of segments that makes a relatively small angle with respect to the applied field. Therefore, a rotational FMR study strongly reflects the symmetry of the lattice and the shape anisotropy of segments [15, 16], which can be systematically varied by fabricating ASI having different segment cross sections and lengths.

FMR spectra can also reveal different types of magnetic ordering, based on the field dispersion of the resonance frequency, df/dH. For example, chiral order (i.e., a sublattice with a flux closure state) is expected to have degenerate FMR modes at zero field. This degeneracy is lifted by applying an external field **H**, such that the resonances of segments with Ising spins aligned parallel (antiparallel) to the field have positive (negative) df/dH [33]. The Fibonacci-distorted square ASI exhibits such a flux closure motif, which is also implicated from step anomalies and plateaus observed in the magnetization M(H) [18].

This paper is organized as follows: First, we describe the geometry of the Fibonacci distortion of a connected KASI. Second, we characterize the branches of FMR modes for the case of Ising saturation, and show how the modes and their bandwidths are modified by Fibonacci distortions of several severities (parameterized by the ratio $r \equiv$ L/S of long (L) and short (S) Fibonacci spacings). Then, we examine field-sweep FMR data, especially in the low-field region, which show that the distortion also causes anomalies in the reversal behavior. Our detailed micromagnetic simulations of the FMR response enable a detailed understanding of these behaviors. The results suggest that FKASI have tunable resonance frequencies and bandwidths suitable for magnonic devices. Moreover, we show the segment reversals can be controlled by the degree of distortion $r$. Finally, we show the low-frequency FMR modes correspond to the resonances near lattice vertices, which we refer to as "localized domain wall modes" (LDW). We show how the aperiodic distortion affects DW modes by changing the shape anisotropy of the vertices, which may find applications in the design of reconfigurable magnonic devices. Technical details, including sample fabrication, FMR measurement techniques, and numerical simulations, are given in an Appendix near the end of the paper.

## Geometry of a FKASI

The KASI can be generated from a periodic honeycomb lattice that has a two-site basis (note the type A and B vertices), as shown in **Fig. 1 (a)**. A Fibonacci distortion is applied to the honeycomb lattice by first replacing the primitive lattice translation vectors, $\boldsymbol{a}$ and $\boldsymbol{b}$ by a chain of "long" or "short" distances corresponding to the "Fibonacci word" **[34]:**

$$S_n = S_{n-1}\ S_{n-2}\ ,$$

where $n \geq 2$, $S_0 \equiv 0$ and $S_1 \equiv 01$. When applying the word to a lattice distortion, 0 corresponds to "long" (L) and 1 corresponds to "short" (S), as shown in **Fig. 1**. The relative lengths between the long and short primitive translation vectors can then be adjusted such that the ratio, $r = L/S$, varies from $r = 1.00$ (**Fig. 2 (a)**) for the undistorted case to $r = 1.62$ (**Fig. 2 (e)**) for the most distorted pattern. This type of distortion can be applied in a very straightforward fashion in the case of square ASI **[18]** where the primitive translation vectors are orthogonal. In contrast, the primitive lattice translation vectors of the honeycomb lattice are not orthogonal, and although the underlying parallelogram lattice is distorted in a unique fashion, the basis site (violet dots in **Fig. 2 (a)**) in the distorted array can be chosen in various ways: we have chosen to place the basis site in the center of a triangle formed by the distorted primitive lattice points of the honeycomb lattice (green dots in **Fig. 2 (a)**). An example of a fabricated sample with $r = 1.62$ is shown in **Fig. 3.**

Note that the 2D Penrose P2 tilings (P2T) are true quasicrystals, and an example of five-fold rotational symmetry that cannot be created by continuous distortion of a 2D Bravais lattice (they are topologically inequivalent). Moreover, the P2T has five mirror planes, any one of which could define the Fibonacci sequence of planar spacings shown in **Fig. 1**. In contrast, the Fibonacci-distorted honeycomb exhibits only one mirror plane and no rotational symmetry, even though it is a continuous distortion of a sixfold Bravais lattice.

## Results and Discussion

Broadband (BB) FMR spectroscopy was performed using a vector network analyzer (VNA); the applied DC magnetic field was in the **x**-direction (shown in **Fig. 3**). We used an end-launch connector to connect a microstripline to the VNA, and placed the samples facing the microstripline ("flip-chip" geometry). Details about the set up are given in the Appendix. We sweep the frequency from 8 GHz to 16 GHz to find higher frequency modes, and from 3 GHz to 9 GHz to find lower frequency modes for a given applied field value. For each sample, we applied a field H = +3000 Oe and extracted $S_{12}$ from the VNA output while sweeping the frequency. Then, we swept the magnetic field from +1000 Oe to -600 Oe, and recorded $S_{12}$ from the VNA output at each field. $S_{12}$ data for +3000 Oe were subtracted from lower-field data for background signal removal. We use the Object Oriented Micromagnetic Framework (OOMMF) to simulate the FMR spectrum for comparison to our experimental data and further analysis. Details about the simulations are given in the Appendix.

### 1. BB FMR Data at H = 1000 Oe

**Figures 4 (a)** and **(b)** show experimental and simulated FMR spectra in an applied magnetic field H = 1000 Oe for distortion ratios, $r$ = 1.0, 1.15, 1.3, 1.45 and 1.62; The distortion creates five groups of segments, labeled I-V according to their easy axis orientation with respect to the applied field (see Roman numerals I-V shown in **Fig. 2 (e)**). We will show that each FMR mode corresponds to a resonant response located within one of these groups of segments: Therefore, we label the corresponding FMR Modes I-V, as shown in **Figs. 4 (a)** and **(b)**.

The spatial designs of patterned magnetic films play a crucial role in tuning the spin wave modes. The Fibonacci distortion of the honeycomb lattice therefore significantly alters the FMR modes observed in both experiment and simulations: We observe frequency shifts and broadening of FMR modes, and new FMR modes emerge as the ratio $r$ increases. For example, we can clearly see Mode II broadening in the experimental data as $r$ increases, as shown in **Fig. 4 (a)**. In the most severe distortion case, $r$ = 1.62, we can see that Mode II splits into Modes II and III. However, the splitting of Mode II can be found in simulations for $r \geqslant 1.3$, as shown in **Fig. 4 (b)**. Moreover, the Mode II and III frequencies shift higher, indicating a reduction of demagnetization field with increasing $r$. Furthermore, we found Mode IV emerges for $r \gtrsim 1.3$ without a noticeable change with increasing $r$, except for a slight frequency reduction, as can be seen in **Fig. 4 (a)**. Mode I does not change with distortion in the simulation; however, we observe a slight broadening in the experimental data. Note that we did not observe Mode V in experiments, although we found this mode in the simulations just below the Mode II frequency, as can be seen in **Fig. 4 (b)**.

Next, we confirm that Modes I-V correspond to resonant response located in Segments I-V, respectively, by comparing the spatial distribution of FMR absorption (mode profile) with simulation results. The mode profiles for $r = 1.62$ are shown in **Figs. 4 (c)-(g)**, where Modes I-V are highly visible inside the bodies of Segments I-V (bulk modes). This behavior is expected, since the higher (lower) frequency modes correspond to segments with an easy axis more aligned (tilted) with respect to the applied field. Note the easy axes of Segments I-V for $r = 1.62$ make angles with the **+x**-direction of 0º, 60º, 47.5º, 7.5º and 78º, respectively.

### 2. Frequency-Field Sweeps and Reversal Behavior

FMR mode behavior in the Ising saturated regime can be well characterized by measuring resonance frequencies as the applied magnetic field is swept from +1000 Oe to -600 Oe. Experimental and simulated frequency-field graphs are shown in **Figs. 5 (a)-(c)** for $r = 1.0$, 1.3, and 1.62, respectively. The modes are labeled I-V in the same manner as in the previous Section, and we found good agreement between experimental and simulation results. All modes have positive df/dH for fields in between 1000 Oe and -300 Oe. Below 500 Oe, for $r = 1.3$ and 1.62, Mode IV disappears. Furthermore, for $r = 1.62$, Modes II and Mode III merge into a single resonance mode, which is consistent with simulation results shown in **Fig. 5**.

Aside from mode dynamics, we find reversal behavior is also affected by increasing $r$. Given our fabricated segment dimensions, it is known that the undistorted KASI does not exhibit a two-step reversal, or that the two fields of the step anomalies are very close together **[6-7]**. However, an unexpected FMR mode appears during reversal for $r = 1.3$ and $r = 1.62$. For $r = 1.0$ and fields down to H = -300 Oe, we could observe two major Modes I and II with positive df/dH. Mode II disappears near H = -300 Oe and the amplitude of Mode I gradually decreases down to -350 Oe (See Appendix **Fig. A2**). We are motivated to define a "reversal event" near -350 Oe, where Mode I-R (we will indeed show that this mode corresponds to the reversal of Segments I) suddenly appears with negative df/dH and gradually increasing absorption intensity. Also, Mode II-R (which we will show corresponds to the reversed Segments II) appears at -400 Oe with negative df/dH. The results for $r = 1$ suggest a large number of segments reverse at -350 Oe, and that there is no apparent two-step anomaly during reversal.

We now discuss the effects of the distortion on modes during reversal. For $r = 1.3$ in **Fig. 5 (b)**, we observe Mode I (Mode II) at applied fields in between H = 1000 Oe and H = -350 Oe (H= -300 Oe), while Mode I is observed with smaller absorption intensity near -350 Oe. We find an unexpected (i.e., an analogous mode is not observed for non-distorted samples) FMR mode in the interval, -300 Oe to -400 Oe, with positive df/dH. This mode is labeled "N", as shown in **Fig. 5 (b)**, both in experiment and simulation. Mode I-R suddenly appeared at H = -300 Oe with negative df/dH. The amplitude of Mode I-R gradually increases with decreasing field.

For $r = 1.62$, we observed Mode I with noticeable amplitude, which persisted down to a field of -350 Oe where Mode I-R suddenly appeared. The amplitude of Mode I-R gradually increases with decreasing field. Another Mode N' appears at H = -300 Oe, and has almost zero df/dH down to H = -400 Oe.  Another Mode II-R appears at -450 Oe.

The FMR experimental data and simulation results in the reversal region (in between H = -300 Oe and H = -400 Oe) indicate the distortion creates the anomalous Modes N and N', and changes the sequence of segment reversal events, which motivated additional investigation of mode profiles in the reversal region. This clarifies the evolution of segment configurations during reversal, and their effect on the internal field and FMR resonance behavior. We show that the abrupt appearance of Mode N in the spectrum correlates with segment reversal events that change the demagnetization field of the sample. For example, **Figs. 6 (a)-(c)** show magnetization configurations for $r = 1.3$ during reversal for -380 Oe, -440 Oe and -460 Oe. The magnetization configurations show that the reversal begins in segments at the boundaries of the FKASI at -380 Oe, followed by reversal of another group of segments at -440 Oe. We can see that all segments are reversed at -460 Oe. Consequently, we have a coexistence regime of reversed and non-reversed segments for fields in the range, – 460 Oe ≤ H ≤ -380 Oe.

Note that, within the reversal regime, some ***chiral states*** (closed loops of magnetization; see **Fig. 6 (b)**) are observed in the simulated magnetization. The loop magnetization is composed of segments with opposite magnetization, which causes the FMR spectrum to exhibit the coexistence of modes with opposite values of df/dH **[8]**. We can identify evidence of chiral states by examining the strongest mode profiles simulated during magnetization reversal, as shown in **Fig. 7** for H = -440 Oe. By comparing **Figs. 7** and **6 (b)**, we can see the mode with frequency 10.2 GHz (Mode N) resides in non-reversed segments, while the mode with frequency 11.7 GHz (Mode I-R) resides in reversed segments. The chiral state should be signaled by splitting of these bulk-mode frequencies (see magnetization texture in **Fig. 6** and Modes I-R and N in **Fig. 5 (b)**).

These observations imply the distortion creates a chiral state during reversal with a significant change in demagnetization field, and indicates spectra will be sensitive to the value of *r*. Evidence for the chiral state is important, since it dominates the long-range-ordered Kagome ground state, making the honeycomb lattice an attractive mesoscopic system that can be directly imaged as a model 2D metamaterial with one or more phase transitions **[5]**. Our present study adds interesting effects of "intermediate disorder" on the chiral state to this list.

### 3. Localized Domain Wall Modes

Domain walls form in the vertices of connected ASI **[14]**, which influences the FMR modes differently compared to the case of disconnected Ising segments. Depending on the shape of the vertex and DW type, peculiar FMR modes can be locally excited within and near vertices. These modes usually have lower resonance frequencies compared with segment bulk modes because of the lower internal fields (higher demagnetization field) present in the vertex region **[21]**. These modes can be useful for designing magnonic devices in which patterned magnetic structures are utilized to gate spin waves in logic and data storage devices **[23]**. FMR mode characterization is therefore an important step towards designing functional magnonic devices. We now show how the distortion affects the vertex shape, and consequently, changes FMR mode characteristics in the vertex region.

Two types of modes related to the vertices are observed. First, vertex center modes (VCM) **[21]** extend throughout and slightly beyond the vertex. Second, LDW modes are more localized in DW at vertices. The VCM and LDW modes are identified in experimental and simulation results shown in **Fig. 5 (a)-(c)** for ratios *r* = 1, 1.3 and 1.62. VCM and LDW modes both have positive df/dH, and the slope for VCM is larger than that for LDW modes.

A VCM was observed experimentally only for *r* = 1.0 with a higher frequency compared to the LDW modes. On the other hand, we experimentally observed LDW modes for all ratios, *r*. Note that LDW modes in the experiment are broader in frequency compared to the VCM, for larger $r$ (See Appendix **Fig. A3**). The simulation results are consistent with the experiment. In **Fig. 5 (a)-(c)**, we can see several LDW modes for $r = 1.0$. As $r$ increases, the number of LDW modes increases, and experiment detects them as a single, broadened mode. We will use the mode profiles to show that an increase in the number of LDW modes is a consequence of the increasing distortion of the vertex shape as $r$ increases. The shape of a vertex depends on the three connecting segments that make $120^\circ$ angles with respect to one another for $r = 1.0$. The Fibonacci distortion changes the angles between segments in vertices and, therefore, forms different vertex types as the honeycomb lattice is distorted (these vertex types are shown in Appendix **Fig. A4**).

We confirmed the VCM and LDW modes reside inside the vertices and DW, respectively. We plot mode profiles for H = 1000 Oe with $r = 1.0$ and $r = 1.62$, in **Figs. 8 (a) - (e)**. The VCM resonance intensity extends throughout the vertex for both values of $r$; whereas the LDW mode intensity is more localized within the DW region. Domain walls for a non-distorted and a distorted vertex (simulated for $r = 1.62$) at 1000 Oe are shown in **Figs. 9 (a)** and **(b).** Comparisons of the LDW mode profile with DW locations show that the LDW mode profiles are highly sensitive to small changes of the locations of the DW.

The behavior of the low-frequency modes can be summarized as follows: (1) VCM exist in non-distorted vertices. Note that the applied field is in the direction of one of the segments' easy axes, which is necessary for VCM to exist **[21]**. (2) VCM do not exist in distorted vertices with applied field in +x-direction, since the easy axes of many segments are no longer in the direction of the applied field. (3) LDW modes exist within vertex DW. Each vertex type can have a specific mode frequency, depending on vertex shape. This suggests that the distortion of the honeycomb lattice can be a useful tool for the design of magnonic crystals.

Finally, we point out a remarkable feature of the "intermediate disorder" of FKASI with potentially interesting applications. Due to the Fibonacci distortion, which is continuously variable, the orientations of some segments and the shape of some vertices are strongly dependent on the ratio $r$. On the other hand, the shapes of other vertices do not significantly change from those for $r = 0$. Moreover, the magnetization textures at the unchanged vertices are not significantly altered, which implies the frequencies of the vertex modes are not significantly altered. For this reason, FKASI spectra possess both changing frequency peaks as well as invariant peaks. From the perspective of applications, this fact can be particularly useful: By changing the *r* parameter, the modes extended along the segments (likely to be important for information delivery) change their frequency and propagation, but at the same time, a subset of VCM maintain their frequencies and can be used as reference control signals as *r* is varied. Hence, the quest for "topological invariant" modes could play a remarkable role when dealing with tunable quasicrystalline ASI. It will be necessary to determine the best design geometry and Fibonacci distortion that would provide easily detectable (intense) VCM/LDW, which is outside the scope of this paper.

## Conclusion and Outlook for Applications

We have demonstrated that the FKASI is an aperiodic array whose FMR modes and segment reversal can be systematically controlled by varying the severity of a Fibonacci lattice distortion of a periodic honeycomb lattice. In particular, multiple modes with controlled resonance frequencies can be designed, and the range of frequencies spanned by a mode can be controlled. Our results show the complex behavior of reversal events, as well as distinct frequencies and frequency-field slopes of FMR modes, can be precisely tuned by varying the severity of the aperiodic lattice distortion. Moreover, the degree of distortion could be a particularly effective parameter for altering the ground state magnetic order and phase transitions among specific sublattices of the FKASI lattice **[14]**. We have detected two types of low-frequency vertex modes (VCM and LDW modes) that are sensitive to the distortion, such that each vertex type generates a LDW mode with a specific frequency. Our results also suggest VCM modes can also be controlled by varying the angle of the applied field.

In summary, Fibonacci distortion serves as a new tool to control FMR modes of patterned magnetic films over a large range frequency, and this type of structure could simplify problems in the design of complicated magnonic systems. Discontinuities in the field-dependent FMR mode frequencies and altered reversal events in distorted samples could be exploited for controlling magnetic switching. One of our principal findings is that the smearing of a resonance peak of the FKASI is not necessarily a negative thing: it is a consequence of the reduction of the segment frequency degeneracy; that is, the distortion creates subsets of segments with different size and orientation, each subset being characterized by a specific frequency and relative intensity. Hence, when magnetic oscillations are associated with binary digits, a distorted ASI can be seen as a multi-signal device, conveying simultaneously more signals in parallel (different frequencies correspond to different oscillation regions), which might have interesting applications as well.

**ACKNOWLEDGMENTS**

Research at the University of Kentucky was supported by the U.S. NSF Grant DMR-1506979, the UK Center for Advanced Materials, the UK Center for Computational Sciences, and the UK Center for Nanoscale Science and Engineering. Research at the Argonne National Laboratory, a U.S. Department of Energy Office of Science User Facility, was supported under Contract No. DE-AC02-06CH11357.

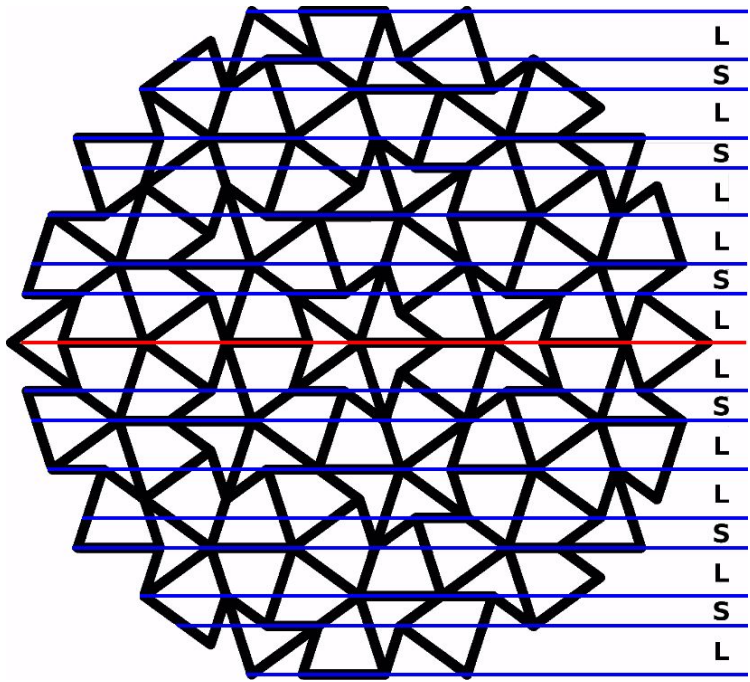

**Figure 1.** Fibonacci spacing in 3rd generation Penrose P2 tiling. L is "long" and S is "short". The ratio of L/S is 1.62. The Fibonacci word in the figure is S4 = 01001010 = LSLLSLSL. Letters on the right margin show the expansion of the Fibonacci word from the center of symmetry in either the up or down directions.

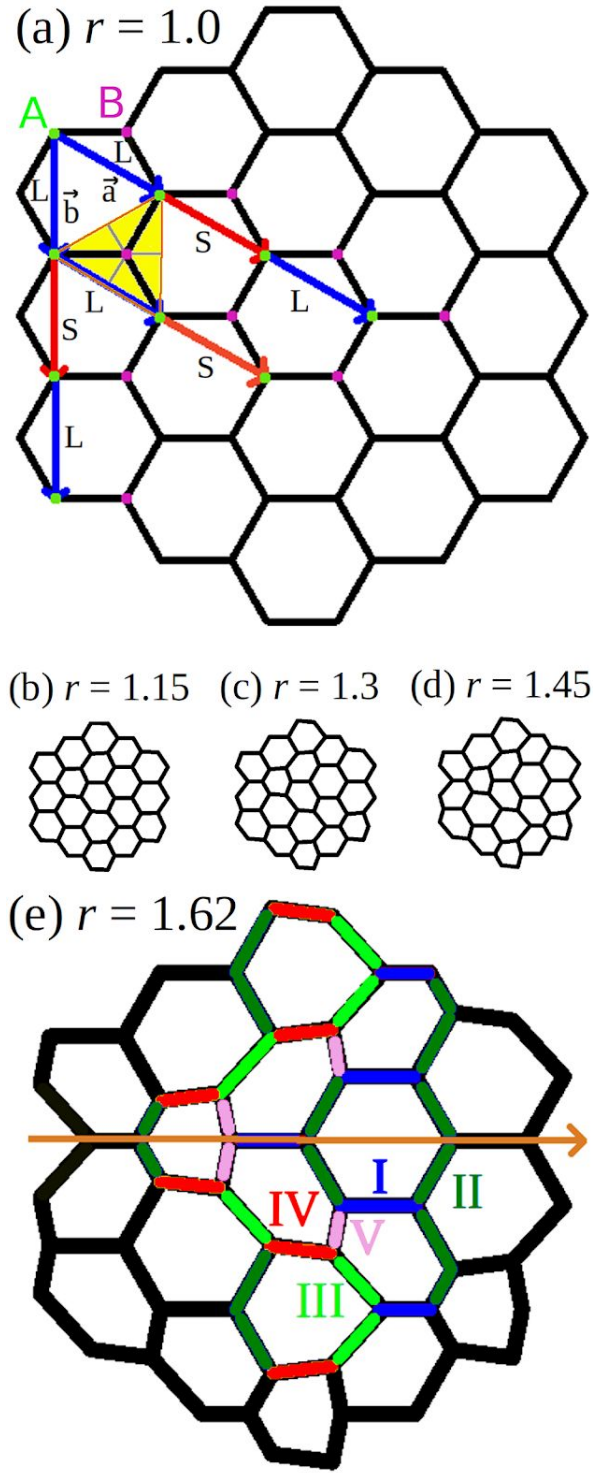

**Figure 2.** Geometry of the KASI distortion. **(a)** Undistorted KASI of third generation ($r = 1$); $\boldsymbol{a}$ and $\boldsymbol{b}$ are the undistorted primitive vectors. Green dots (type-A lattice point) are primitive lattice points, and violet dots (type-B lattice point) define the basis positions. **(b)-(d)**: In the distorted lattice, the length of the primitive vectors is either long (L with blue color) or short (S with red

color), according to the Fibonacci sequence. Distorted ASI are shown for $r =$ L/S $= 1.15$, $r = 1.3$ and $r = 1.45$. **(f)**: Distorted ASI for $r = 1.62$. Different segment types are shown in different colors and labelled by Roman numerals. For example, the blue segments aligned along the **x-**axis are assigned the numeral I. We can see a sixfold rotational symmetry of the undistorted KASI is reduced to one mirror plane that is indicated by the orange vector along the **x-**axis.

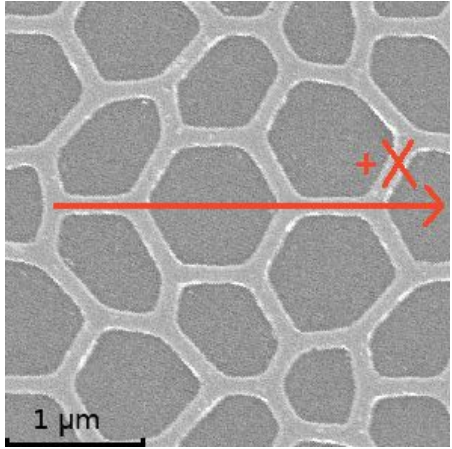

**Figure 3.** SEM image of the FKASI for $r = 1.62$ patterned on a quartz substrate. The orange line denotes the **+x-**direction and the location of a mirror plane. The width of the segments is w ~ 140 nm (note the 1-micron scale bar).

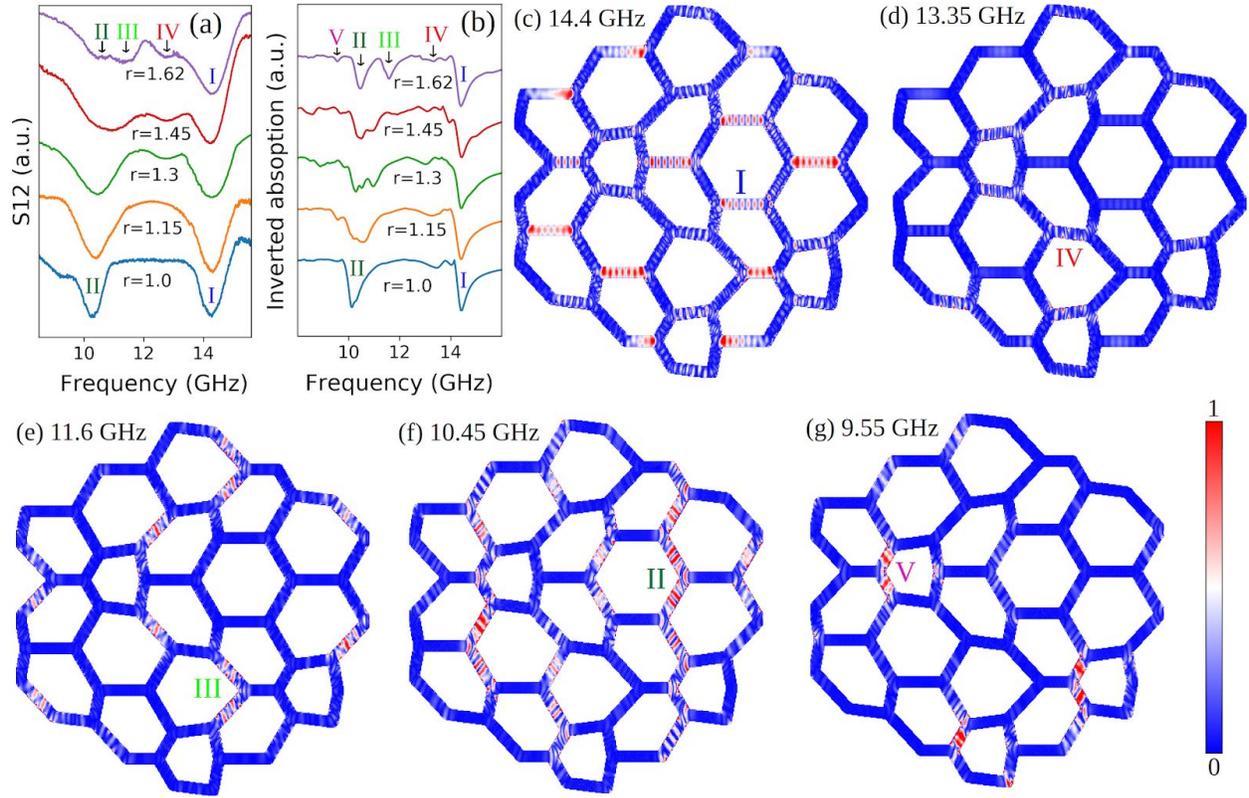

**Figure 4.** FMR results for applied field H = 1000 Oe for different distortion ratios $r$. **(a)** $S_{12}$ (transmission from VNA port 1 to port 2) as a function of frequency for each value of $r$. **(b)** Numerical simulations of the absorption spectrum as a function of frequency for each value of $r$. Note that modes are labeled for $r = 1.0$ and 1.62. The mode labeling is consistent with corresponding segments labeled I-V. **(c)-(g)** Mode profiles for $r = 1.62$ for five peaks of the absorption spectrum. Segments in (c)-(g) are labeled corresponding to the mode labels. The color bar represents absorption intensity, where red is maximum absorption intensity and blue is zero absorption intensity.

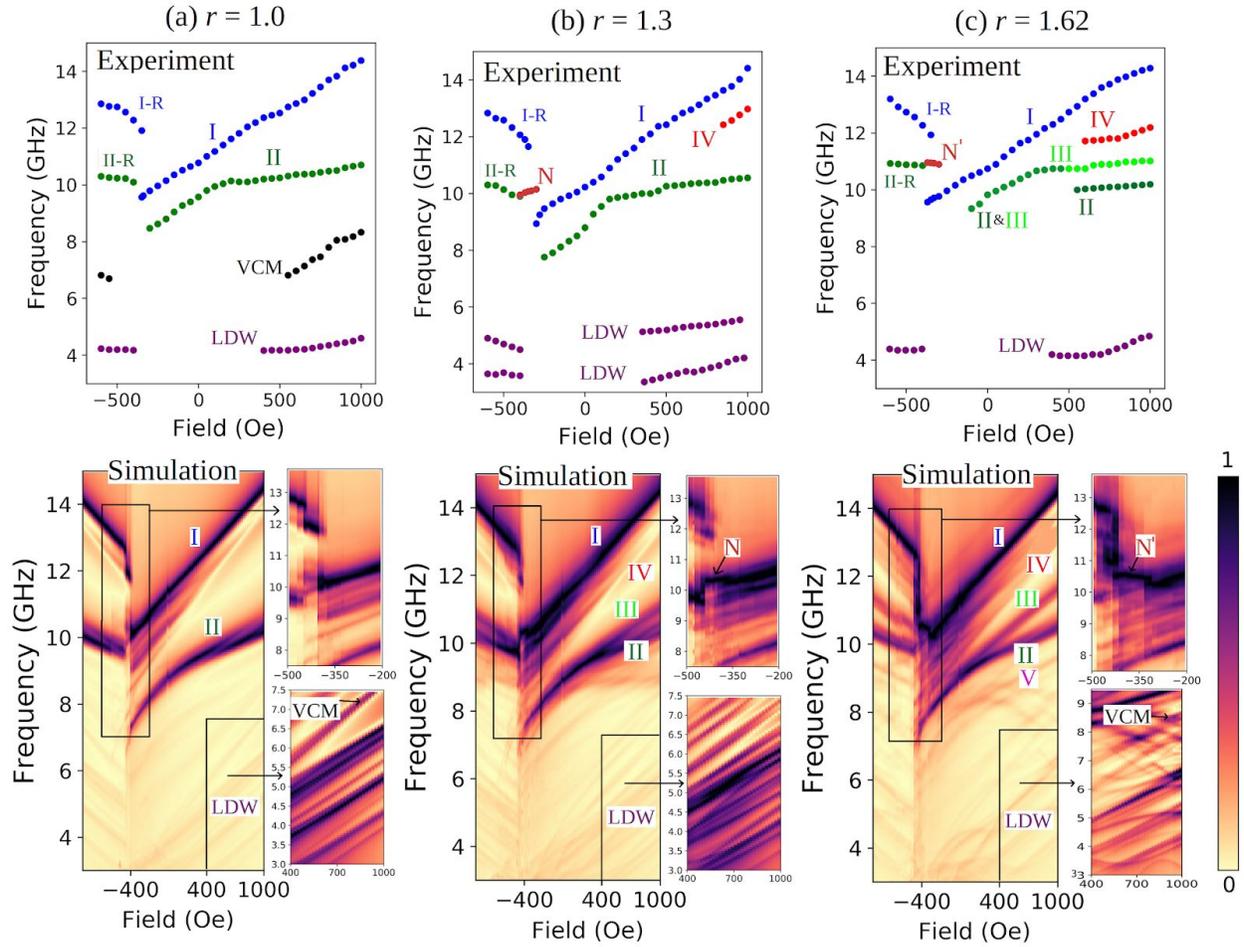

**Figure 5.** Experimental (top panels) and simulation (bottom panels) results for field vs. frequency for samples with the following distortions: **(a)** $r = 1.0$, **(b)** $r = 1.3$ and **(c)** $r = 1.62$. High frequency modes are labelled I, II, II, IV and V, and assigned colors. Modes labeled I-R and II-R are created after reversal. The color bar defines 1 (dark) for maximum absorption intensity and 0 (light) for no absorption.

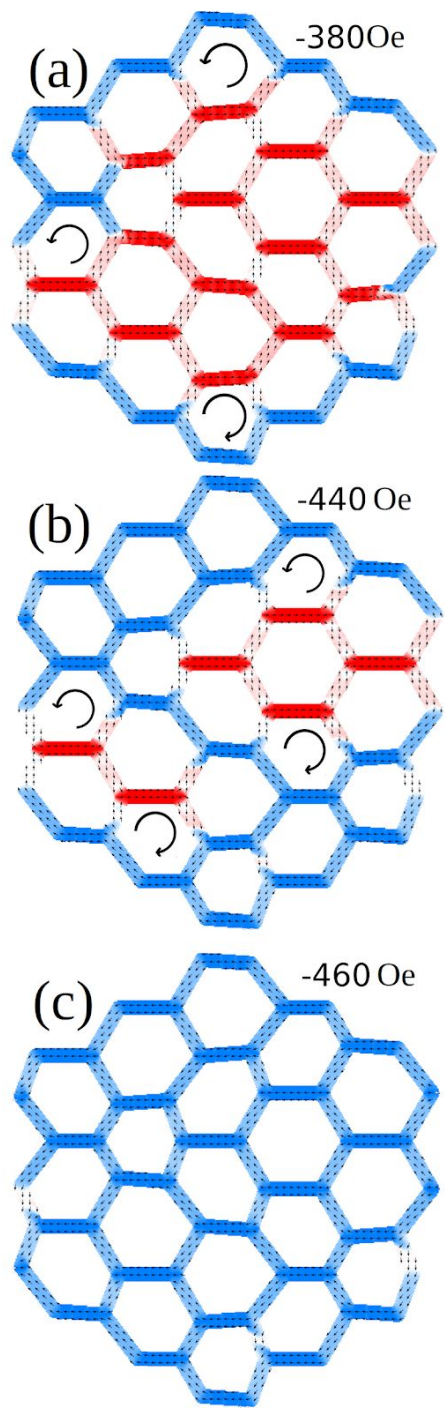

**Figure 6.** Simulated magnetization textures for three values of applied field **(a)** H =-390 Oe, **(b)** H = -440 Oe and **(c)** H = -460 Oe, during a sweep from +1000 Oe to -600 Oe. Magnetization reversal of several segments begins at -390 Oe followed by additional segment reversals at -440 Oe. All segments are reversed at -460 Oe. Blue and red indicate reversed and non-reversed regions, respectively. Small arrows indicate magnetization direction; curved arrows indicate

chiral cells.

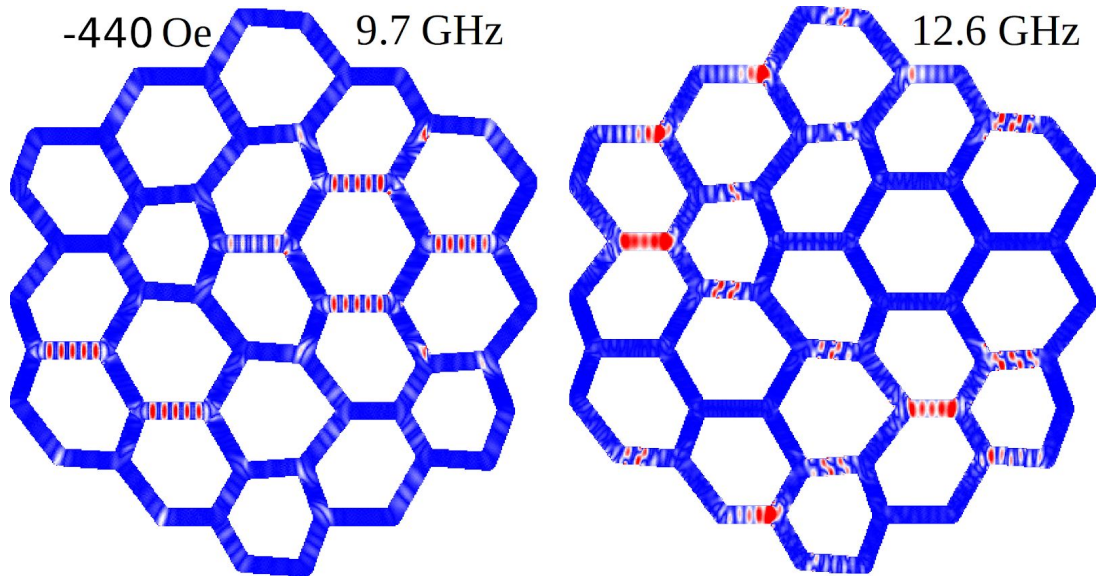

**Figure 7.** Mode profiles at H = -440 Oe for *r* = 1.3. The mode profile for frequency 9.7 GHz (right) exhibits a resonance of non-reversed segments, and the mode profile at 12.6 GHz (left) exhibits a resonance of reversed segments. The color bar represents absorption intensity, where red is maximum intensity and blue is zero intensity.

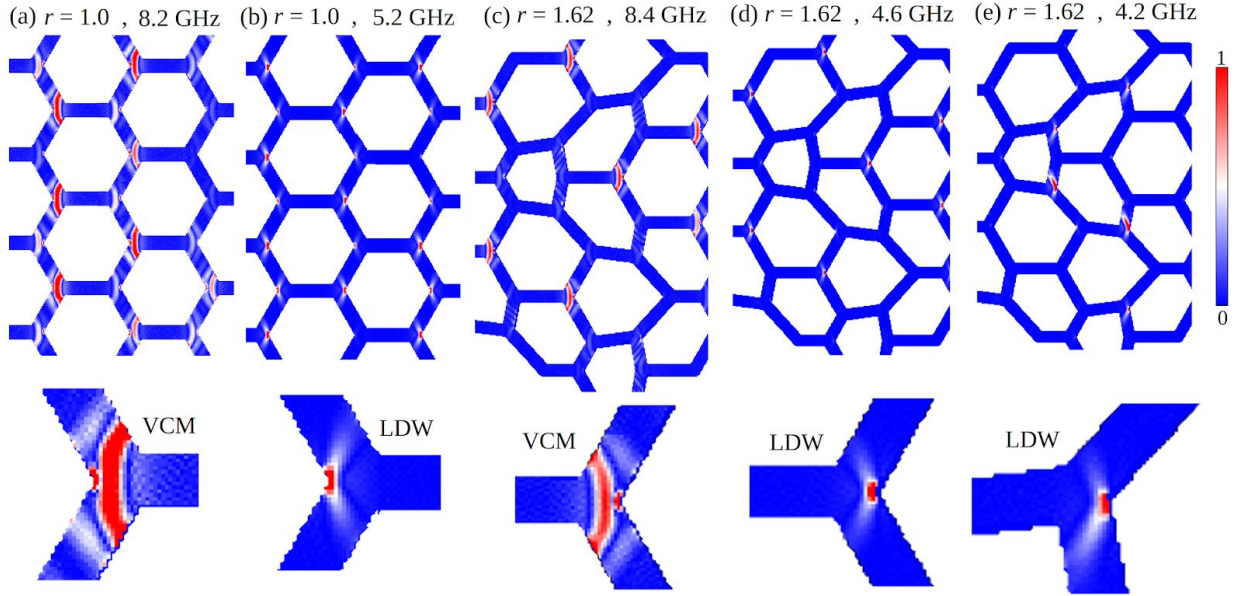

**Figure 8.** Mode profiles for LDW modes. **(a)** and **(b)** show the mode profiles for an applied field H = 1000 Oe for *r* = 1.0 at 8.2 GHz (VCM) and 5.2 GHz (LDW mode), respectively. **(c)-(e)** shows mode profiles for *r* = 1.62 at 8.4 GHz (VCM), 4.6 GHz (LDW) and 4.2 GHz (LDW), respectively. **(d)** and **(e)** show that two different vertex types resonate at different frequencies.

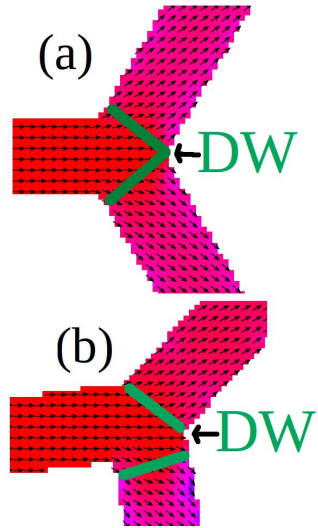

**Figure 9.** **(a)** Magnetization texture of a representative vertex in KASI for r = 1. **(b)** The corresponding texture for a distortion ratio *r* = 1.62. DW are shown by green lines. The applied field is H = 1000 Oe, oriented horizontally to the right.

# Appendix:

## A. Experimental Details

### A.1. Sample Fabrication

Samples were patterned on a quartz ($SiO_2$) substrate. The substrates were cleaned by sonication in acetone and IPA for 5 minutes each, then rinsed in DI water and blown dry with compressed $N_2$. Any remaining organic solvents were removed from the substrates by plasma $O_2$ cleaning. The substrates were then spin-coated with a bilayer of PMMA A4, 495K and 950K resist. Before patterning, a 7-nm-thick layer of Au was sputtered on the bilayer of PMMA A4 to reduce any charging caused by the insulating nature of the quartz substrates. The PMMA was then exposed using electron beam lithography. The Au film was then removed with a standard Au etch solution. The exposed PMMA A4 bilayer was then developed in a solution of ethyl alcohol and DI Water. Samples were then loaded into an electron beam evaporator system. A vacuum of approximately $5.0 \times 10^{-7}$ to $8 \times 10^{-7}$ Torr was achieved before deposition of permalloy at a rate of 0.03 nm/s to the desired thickness of 25 nm. Al was then deposited at a rate of 0.02 nm/s to a desired thickness of 1.5 nm to passivate the permalloy to oxidation. Thickness and deposition rate were estimated with a crystal thickness monitor placed inside the electron beam evaporator chamber. Samples were then developed in *Microposit 1165* for the lift-off procedure.

A final check of pattern quality, location and segment widths was done using SEM imaging in a *Raith eLine Electron Beam Lithography System*. **Figure 2** shows an SEM image of a KASI sample with $r = 1.62$. The patterned sample was a hexagon of approximate $43 \times 43$ um dimension, repeated in a grid of $4 \times 80$ hexagons with a lattice spacing of 100 um. The undistorted hexagonal pattern has a segment length l = 500 um and width W ≈ 140 nm.

### A.2. FMR Methods

The experimental FMR setup includes an electromagnet, a vector network analyser (VNA), and a microstripline. A sample film-on-substrate is placed as a flip-chip on the microstripline, which is composed of a signal line, substrate and ground plane, as shown in **Fig. 3**. The substrate is a 100-micron-thick, industrial-grade Kapton film that is attached using a silicone adhesive to a copper plate that functions as a RF ground plane. The signal line is designed with four parallel conducting strips of 20-micron width and 8-mm length. The center-to-center distance between these strips is 100 micron, and they join at both ends to a conducting strip of 320-micron width (shown in **Fig. 3**, top view). We used standard photolithography to fabricate the signal line. First, we sputtered 200 nm of copper on the Kapton film. Then, we spin-coated the copper-coated Kapton film with S-1813 photoresist, followed by exposure to a high-power laser source under the photomask. The film was then developed using a S-1813 developer. Finally, the redundant copper was etched using diluted ferric chloride, and the remaining photoresist was removed using acetone. The patterned sample was placed as a flip-chip such that magnetic material was aligned with the 20-micron stripes to obtain optimal coupling to RF signals, as shown in **Fig. 3** (top view).

## A. 3. Numerical Simulations

Micromagnetic simulations were performed using the Object Oriented Micromagnetic Framework (OOMMF). In OOMMF, we used a 10-nm-by-10-nm mesh cell size with a 25-nm film thickness. The film material was Permalloy with a saturation magnetization $M_S = 800$ kA/m and exchange stiffness constant $A = 1.3 \times 10^{-12}$ J/m [35]**.** To find the FMR absorption spectrum, we first initialized the magnetization **M** to be uniformly saturated along the applied DC field (**H**) in +**x**-direction. We sweep the field from +1000 Oe to -900 Oe with 20 Oe steps however 5 Oe steps was used for the field range of -200 Oe to -500 Oe (around reversal fields). The magnetization vectors for each cell are recorded. Then, we applied an out-of-plane (**z**-direction) magnetic field of magnitude 10 Oe in a direction perpendicular to H for 20 picoseconds. We recorded the magnetization vectors 1000 times in 20 picoseconds time steps. We used a fast Fourier transform (FFT) to find the spectrum for each film cell. The cutoff frequency is 25 GHz and frequency resolution is 25 MHz. The total absorption spectrum corresponding to the BB FMR data was found by averaging the spectrum of all cells. Also, we found the spatial distribution of the FMR absorption by color plotting of the spectrum in each cell at the desired frequency.

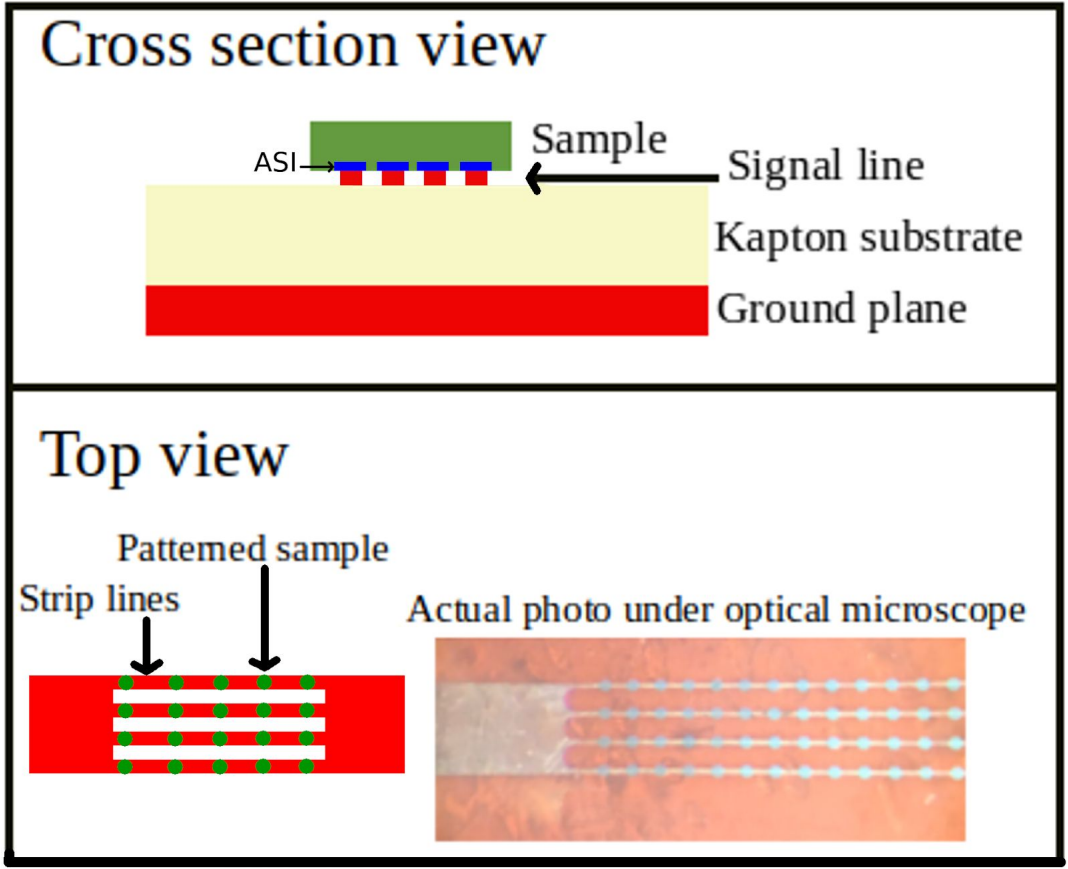

**Figure A1.** Set up for BB FMR measurements. The top figure shows a cross-sectional view. Green denotes the sample. and placed as a flip-chip on the red microstripline deposited on the Kapton substrate, which is attached on the solid copper plane using silicone adhesive. The bottom figure shows the top view of the setup. The red on the lower-left image is the microstripline, and green dots represent arrays of ASI. The bottom right shows an optical image of an ASI sample patterned on a quartz substrate, which is placed as a flip-chip on the microstripline and aligned with the narrow, 20-micron strips.

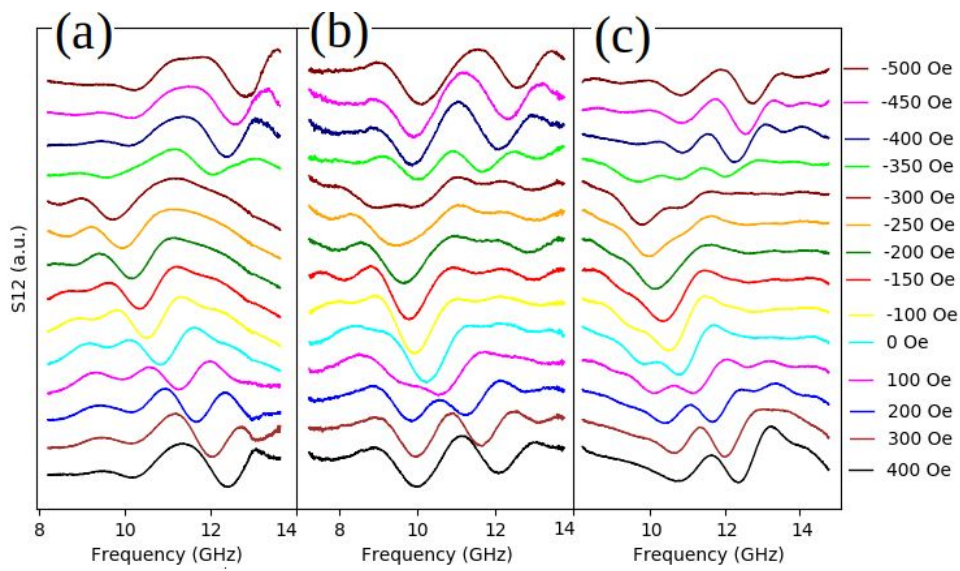

**Figure A2.** S12 as a function of frequency for applied fields -500 Oe ≤ H ≤ 400 Oe for (a) *r* = 1.0 (b), *r* = 1.3 and (c) *r* = 1.62.

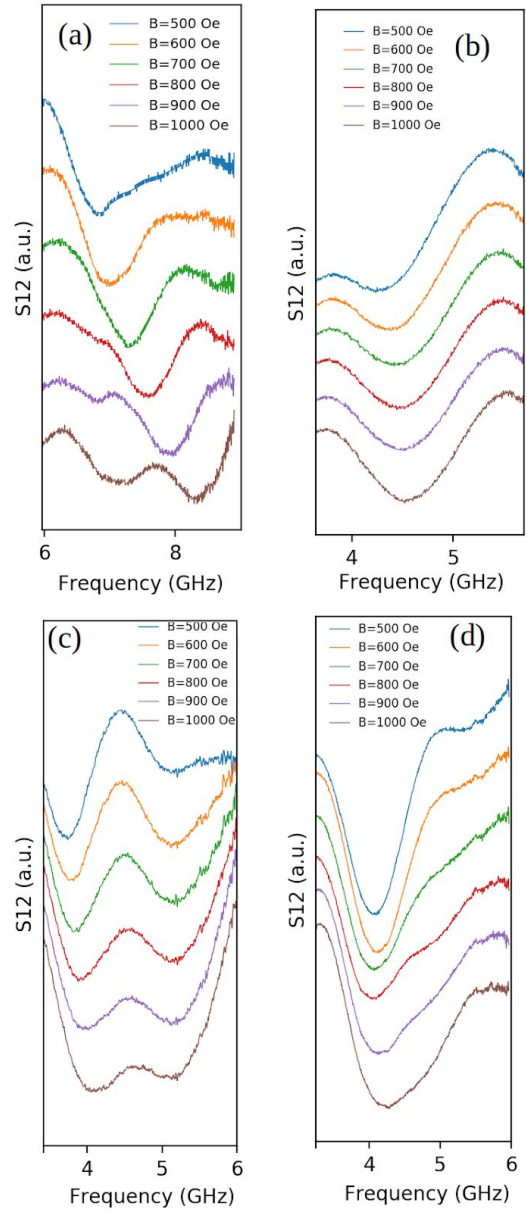

**Figure A3.** S12 as a function of frequency for applied fields 1000 Oe ≤ H ≤ 500 Oe for (a), (b) $r$ = 1.0 (c), $r$ = 1.3 and (d) $r$ = 1.62.

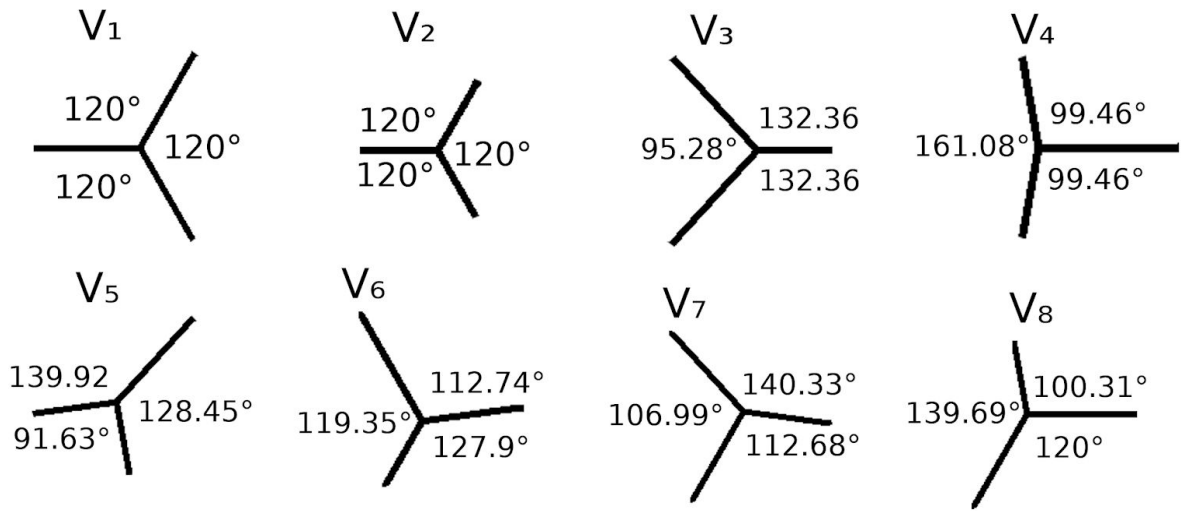

**Figure A4.** Different types of vertices created by the Fibonacci distortion. Angles are shown for *r* = 1.62.